\documentclass[twocolumn,amsmath,amssymb,nofootinbib]
{revtex4}
\usepackage{amsmath}

\usepackage{physics}
\usepackage{subfigure}
\usepackage[T1]{fontenc}
\usepackage[utf8]{inputenc}
\usepackage{graphicx,dcolumn,bm,color,latexsym,amssymb}
\usepackage{import}
\usepackage{xifthen}
\usepackage{pdfpages}
\usepackage{transparent}
\usepackage{verbatim}
\usepackage{hyperref}

\usepackage{hyperref}
\hypersetup{
    colorlinks=true,
    linkcolor=blue,    % Color of internal links
    citecolor=blue,    % Color of citation links
    urlcolor=blue       % Color of external links
}

\definecolor{Blue}{rgb}{0.0,0.0,1}
\definecolor{Red}{rgb}{1,0.0,0.0}
\definecolor{Green}{rgb}{0,0.5,0.0}

\begin{document}
%\title{Effects of the kinetic energy in heat for driven systems: active and passive case.}
\title{Effects of kinetic energy on heat fluctuations of passive\\ and active overdamped driven particles}

\author{Pedro V. Paraguass\'{u}}
 \email{paraguassu@esp.puc-rio.br}
\affiliation{Departamento de F\'{i}sica, Pontif\'{i}cia Universidade Cat\'{o}lica\\ 22452-970, Rio de Janeiro, Brazil}

\author{Rui Aquino}
\affiliation{ICTP--South American Institute for Fundamental Research - Instituto de F\'isica Te\'orica da UNESP, \\
Rua Dr.\ Bento Teobaldo Ferraz 271, 01140-070 S\~ao Paulo, Brazil}

\author{Pablo de Castro}
\affiliation{ICTP--South American Institute for Fundamental Research - Instituto de F\'isica Te\'orica da UNESP, \\
Rua Dr.\ Bento Teobaldo Ferraz 271, 01140-070 S\~ao Paulo, Brazil}
%\date{December 2023}

\begin{abstract}
To describe the spatial trajectory of an overdamped Brownian particle, inertial effects can be neglected.
%because the ratio of mass to damping coefficient gives an extremely small time scale. 
Yet, at the energetic level of stochastic thermodynamics, changes in kinetic energy must be considered to accurately predict the  heat exchanged with the thermal bath. On the other hand, in the presence of external driving forces, one would expect the effects of kinetic energy fluctuations to be reduced, as thermal noise becomes comparatively less relevant. Here, we investigate the competition between the kinetic energy and the external work contributions to the heat statistics of passive and active overdamped Brownian particles subject to external driving forces. We find that kinetic energy effects cause fluctuations in the exchanged heat to become non-Gaussian. To evaluate the relevance of these effects, we compute the excess kurtosis and the Pearson correlation. For fixed parameter values adapted from experiments, we identify a crossover transition from a regime in which the stochastic heat of overdamped particles is dominated by external work, where kinetic energy changes can be safely ignored, to a regime dominated by kinetic energy effects. Our results also provide a quantitative analytical way to assess how deep into a particular regime the system is.
%say that transition can be obtained though simple analysis
%We also show that, although a simpler analysis of the motion equation predicts nearly the same transition driving force, our results also provide a quantitative way to assess how smooth is the transition.
\end{abstract}

\maketitle

\section{Introduction}
When dealing with Brownian particles, fluctuations in energy and energy transfer arise due to the inherent stochastic nature of Brownian motion \cite{tome2015stochastic}. Such energetic fluctuations are interpreted and studied within the framework of Stochastic Thermodynamics \cite{peliti2021stochastic}. Employing this framework allows us to explore the exchange of heat energy between the particle and its environment. Recently, it has been demonstrated that changes in kinetic energy contribute significantly to the heat exchange in \textit{overdamped} particles \cite{paraguassu2022effects}. This result may seem unexpected since, for an overdamped Brownian particle, inertia can indeed be disregarded in the equation of motion describing spatial trajectories since the mass-to-damping coefficient ratio gives a tiny time scale. Such contribution of kinetic energy to heat in overdamped systems remained obscured due to the common, albeit incorrect, neglect of mass-related effects at the \textit{energetic} level \cite{paraguassu2022effects,arold2018heat}. The realization that kinetic energy effects can be quantitatively important even for overdamped systems calls for a reexamination of previously obtained results. 

In this context, a central question is how external driving forces might significantly diminish the impact of kinetic energy fluctuations as thermal noise becomes comparatively less significant. Since the stochastic spatial trajectories themselves fluctuate less in the presence of deterministic external driving forces, our working hypothesis is that such reduction of stochasticity is inherited into the energetic level and that this potentially makes kinetic energy effects less relevant. To which extent this occurs is an open question that we address here. In this work, we study overdamped driven particles and investigate the statistical properties of particle-bath heat exchange with and without changes in kinetic energy. The simple driving forces examined here introduce a bias in the particle's motion which indeed changes the nature of the heat fluctuations. Such driven systems are particularly interesting as they serve as prototypes for work-to-work converter engines \cite{filho2022thermodynamics, mamede2022obtaining}. These driving forces are commonly realized using simply gravity or electrodes in levitated particle experiments \cite{kremer2024all, kremer2024perturbative}. 
%In the active case also considered here, self-propulsion can generate also a stochastic driving \cite{bechinger2016active}.

%In a system where energy contributions arise from both kinetic energy and work generated by an external driving force, the relevance of kinetic energy becomes a critical question. Here, we investigate a driven system where an external force acts directly on the particle. Such systems are used as prototypes for work-to-work engines \cite{}, and this type of external force can be experimentally implemented using electrodes in levitated particle experiments \cite{}.

Earlier predictions for such overdamped driven systems considered \cite{singh2008onsager,saha2014work} that heat $Q$ would be equal to work $W$, implying that the heat exchanged in a fixed time window would be a Gaussian random variable due to the absence of kinetic energy. However, as we shall see, the necessity of incorporating changes in kinetic energy into the heat expression for overdamped particles means that heat no longer follows a Gaussian behavior since \(Q \neq W\). We derive and investigate these non-Gaussian fluctuations of heat by systematically employing the corrected heat expression, computing its characteristic function and moments, and analyzing the emergent non-Gaussian behavior through two different quantities: the excess kurtosis and the Pearson correlation coefficient. We find that, for depending on the parameters, the change in kinetic energy not only cannot be neglected but also is the most significant contribution to the statistical fluctuations of the heat. As driving forces are increased or temperature decreased, the system smoothly undergoes a crossover transition to a regime where kinetic energy effects can again be ignored. More importantly, our results provide a way to assess how smooth is the transition, or how deep a system is into one regime or the other.

We also extend our results to Brownian particles subject to an additional active noise \cite{bonilla2019active}. In this case, the equation of motion models the movement of a self-propelled particle such as an organism or the movement of a passive particle in an active bath consisting of self-propelled particles that hit our probe particle. In both cases, our particle becomes active and thus moves in a persistent fashion \cite{bechinger2016active,de2023sequential}. We find that the type of activity considered here qualitatively enhances the fluctuations of the kinetic energy, making this term more relevant. %Quantitatively, for the experimental parameters studied here, activity adds a small contribution to the heat variance. 

This paper is organized as follows. In Section \ref{modelandthermo}, we define the model and its thermodynamics. In particular, we define heat as the exchange of energy between the particle and the passive thermal bath, for both the active and passive cases, and include the effect of kinetic energy. In Section \ref{distribution}, we compute the heat fluctuations through kinetic energy and work, revealing a non-Gaussian distribution. The next step is to investigate the non-Gaussian statistics in Section \ref{nongaussian} based on experimental values for both passive and active cases. We calculate the excess kurtosis and Pearson correlation coefficient that allows us to identify the crossover transition to the kinetic energy dominated regime.
%, finding that kinetic energy could be relevant and becomes increasingly so as the temperature of the passive thermal bath is raised. 
Finally, we conclude in Section \ref{discussion} with a summary and a discussion of our results.

\section{Model and Thermodynamics}\label{modelandthermo}

We are interested in evaluating the contribution of kinetic energy to the heat fluctuations of overdamped driven particles, first in a passive case and then in an active one. In both scenarios, we define heat as the exchange of energy between the particle and the \textit{passive} thermal bath. 

\subsection{Passive particle}
Consider an overdamped passive particle driven by a deterministic force, e.g., a silica particle subject to a force due an electric field \cite{dassanayake2000structure}. The dynamics of the particle obeys the following vector equation of motion in arbitrary spatial dimensionality $d$:
\begin{equation}
    \gamma \boldsymbol{v}(t) = \boldsymbol{f}(t) + \boldsymbol{\eta}(t),\label{langevinp}
\end{equation}
where $\gamma$ is the damping coefficient, $\boldsymbol{v}(t)$ is the velocity and $\boldsymbol{f}(t)$ is the deterministic driving force performing work on the particle. Also, $\boldsymbol{\eta}(t)$ is a stochastic force---due to a passive thermal bath---with zero mean and correlation
\begin{equation}
    \langle \eta_i(t)\eta_j(t^\prime) \rangle = 2\gamma k_B T  \delta_{ij}\delta(t-t'),\label{passivenoise} 
\end{equation}
where $\eta_i(t)$ is the $i$-th spatial component of $\boldsymbol{\eta}(t)$.
This noise is the result of collisions between the molecules of the passive thermal bath and the particle over a fast time scale. For the sake of generality, we allow the driving force $\boldsymbol{f}(t)$ to be time-dependent but, for simplicity, we will switch to a constant driving force when a concrete form of the driving force is required.

The work that the deterministic force does on the particle over a process between time $t=0$ and an arbitrary $t=\tau$ is
\begin{equation}
    W[\boldsymbol{v}] = - \int_0^\tau \boldsymbol{f}(t) \cdot \boldsymbol{v}(t) dt.
    \label{workdef}
\end{equation}
In our scenario, the first law of thermodynamics states that, for instance, an increase in the kinetic energy of the particle comes from the combination of heat gained from the environment (through collisions with the particles composing the passive thermal bath) plus an additional energy coming from the work done by the external force on the particle. Therefore, as discussed in Ref.\ \cite{paraguassu2022effects}, the heat exchanged in that process is given by
\begin{equation}
    Q[\boldsymbol{v}] = \Delta K + W[\boldsymbol{v}],
\end{equation}
where $\Delta K = m(v_\tau^2-v_0^2)/2$ is the kinetic energy change, with $v_0$ and $v_\tau$ denoting the initial and the final velocities in the process. As per usual, for $Q>0$ ($Q<0$) the particle has absorbed (released) energy from (into) the passive thermal bath.

Notice that this expression for the heat $Q$ involves the change in kinetic energy $\Delta K$. In constrat, the seminal definition of heat by Sekimoto for overdamped particles \cite{sekimoto2010stochastic} does not include the kinetic energy change. One simple way to describe an argument that has been used for neglecting the kinetic energy term is this: since mass does not appear in the overdamped equation of motion, terms involving mass such as kinetic energy can also be eliminated from the heat definition; alternatively, one can say that, in the overdamped regime where inertia can be neglected, the velocity is a fast variable. However, by making sure that we eliminate only terms where the mass-to-damping coefficient ratio is much smaller than other relevant time scales of the process, it has been noted that kinetic energy effects survive and are strong \cite{paraguassu2022effects}. In fact, as demonstrated in Ref.\ \cite{paraguassu2022effects}, and generalized in Appendix \ref{A} for the active case, the kinetic energy contribution arises naturally if one starts from the underdamped Langevin equation together with Sekimoto's definition and then takes the overdamped limit appropriately.

\subsection{Active particle}
For active particles, movement is also influenced by an active stochastic force which stems from local conversion of fuel sources into propulsion \cite{bechinger2016active}. As a result, activity produces a persistent, i.e., time-correlated dynamics. Classical examples include self-propelled organisms such as fish and cells, or a passive particle diffusing in a ``sea'' of self-propelled organisms, which acts as an active bath, plus a passive bath. The equation of motion now becomes
\begin{equation}
    \gamma \boldsymbol{v}(t) = \boldsymbol{f}(t)+\boldsymbol{f}_a(t) + \boldsymbol{\eta}(t),\label{activemodel}
\end{equation}
where $\boldsymbol{f}_a(t)$ is the active noise. This is chosen to be an Ornstein-Uhlenbeck variable with zero mean and correlation
\begin{equation}
    \langle \boldsymbol{f}_a(t)\cdot\boldsymbol{f}_a(t^\prime) \rangle = v_a^2\gamma^2\exp\left(-|t-t'|/\tau_a\right).
    \label{activenoise}
\end{equation}
The model contains two activity parameters that control the statistics of the fluctuating active force $\boldsymbol{f}_a$: the active velocity scale $v_a$, which is proportional to the standard deviation of the active force, and a persistence or correlation time $\tau_a$.
%The parameter $n$ depends on the spatial dimensionality of the problem. 
This choice of an active noise whose \textit{magnitude} oscillates and has zero mean corresponds to the active Ornstein-Uhlenbeck particle (AOUP) model \cite{bonilla2019active,shankar2018hidden}. The AOUP model can be used to accurately describe a passive particle diffusing in a dilute active bath where ``friction'' due to collisions with the active particles is absent \cite{maggi2014generalized}. Working with the AOUP model is an analytically convenient choice. Furthermore, although truly self-propelled particles are usually modeled using the active Brownian particle (ABP) model \cite{rojas2023wetting,rojas2023mixtures}, the AOUP model can also be used as an approximation for self-propelled particles such as epithelial cells \cite{deforet2014emergence}.

In Ref.\ \cite{dabelow2019irreversibility}, two heat definitions for an AOUP are discussed. They are called ``active-bath'' and ``self-propulsion'' scenarios. In the ``active-bath'' scenario, active forces come from a surrounding bath of active particles and are interpreted as just an additional external force. In the ``self-propulsion'' scenario, the activity model is understood as a ``coarse-grained'' description that does not allow us to assess how much energy the conversion of fuel into self-propulsion dissipates. What we can calculate is the dissipation associated with deviations from the self-propulsion trajectory. These deviations result in a friction $-\gamma\boldsymbol{v} + \boldsymbol{f}_a$ with the passive bath and with the active self-propulsion mechanism. In this case, one calculates the heat transfer between particle and passive bath that does \textit{not} come from active sources. Notice that, indeed, the active contribution to movement is subtracted from the friction. Since we have not really specified which of these two scenarios we are working on, we choose the latter heat definition for analytical convenience. Furthermore, we understand that heat defined in this way can be relevant in both scenarios despite how they were called in Ref.\ \cite{dabelow2019irreversibility}. 

Accordingly, the heat in the active case is defined as
\begin{equation}
    Q[\boldsymbol{v}] = \Delta K + \int_0^\tau \left(\gamma (\boldsymbol{v}- \boldsymbol{f}_a\gamma^{-1}) + \boldsymbol{\eta}(t) \right) \cdot \left(\boldsymbol{v}- \boldsymbol{f}_a\gamma^{-1}\right) dt,
\end{equation}
where the second term on the right-hand side corresponds to the heat expression given by \cite{dabelow2019irreversibility} and we included by hand, for now, the contribution of the change in kinetic energy. By using the AOUP equation of motion, Eq.~\eqref{activemodel}, we find
\begin{equation}
    Q[\boldsymbol{v}] = \Delta K -\int_0^\tau \boldsymbol{f}(t) \cdot \left(\boldsymbol{v}- \boldsymbol{f}_a\gamma^{-1}\right) dt.\label{activeheat}
\end{equation}
Here, besides the kinetic energy contribution, we have the usual work and a new term due to the active noise. In fact, this new contribution is a form of heat as pointed out in \cite{dabelow2019irreversibility}, and arises due the disordered energy exchanged between the particle and the active fluctuations. We will call it active heat. Importantly, this expression can also be derived by starting from the underdamped equation of motion without explicitly including the kinetic energy, which in turn appears naturally due to inertial effects at the energetic level. We provide this derivation in Appendix \ref{A}.

\section{Fluctuations of energy exchanges}\label{distribution}

As we shall see, due to the contribution of kinetic energy changes, the exchanged heat $Q$ is not a Gaussian variable in either the passive or active case. The fluctuations of the exchanged heat can be completely characterized by the characteristic function $Z(\lambda)$, which can be used to compute all the statistical moments. For a generic stochastic variable $X$, its moments can be calculated through
\begin{equation}
\langle X^n \rangle =  i^n\frac{\partial^n Z(\lambda)}{\partial \lambda^n}\bigg|_{\lambda\rightarrow 0}.
\end{equation}
Although our overdamped Brownian particles can move in three dimensions, many experimental setups constrain them to two dimensions, particularly in active matter \cite{bechinger2016active}. Therefore, from now on, we will consider $d=2$ dimensions. In this section, we will construct the characteristic function of heat based on three quantities: the distributions of velocities and the characteristic functions of the kinetic energy as well as of the work.

\subsection{Distribution of velocities}

First, we will need to know the stationary distribution of the initial (or final) velocities of the overdamped particle. In the passive case without external driving forces, the distribution is just the Maxwell-Boltzmann distribution:
\begin{equation}
    P(\boldsymbol v_0) = \left(\frac{\beta m }{2\pi}\right)^{d/2}\exp\left(-\frac{\beta m}{2}\boldsymbol v_0^2\right)
\end{equation}
where $\beta\equiv1/k_{B}T$ and $d=2$. However, since we are applying an external driving, we have a mean velocity proportional to the applied force. Therefore the distribution needs to take this mean into account and becomes
\begin{equation}
    P(\boldsymbol v_0) = \left(\frac{\beta m }{2\pi}\right)^{d/2}\exp\left(-\frac{\beta m}{2}\left(\boldsymbol v_0-\boldsymbol f_0/\gamma\right)^2\right).
\end{equation}
Here, we assumed that the external driving is a constant force. This simple case will facilitate our investigation regarding the relevance of kinetic energy effects. 
We now have a non-equilibrium steady state distribution \cite{taniguchi2008inertial, hatano2001steady}.

\begin{figure}[htbp]
    \centering
    \includegraphics[width=0.9\columnwidth]{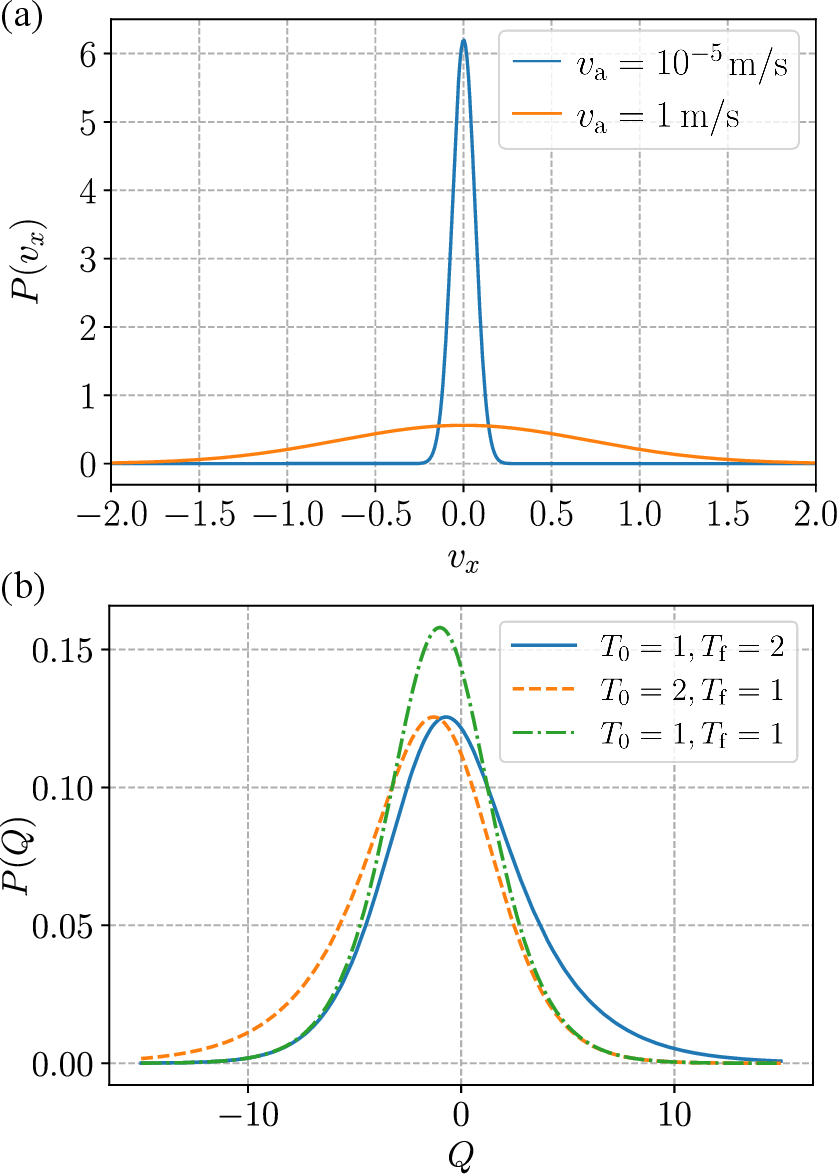}
    \caption{(a) Distribution of velocities [Eq.~\eqref{steadyactive}] along the $x$ direction for an AOUP subject to thermal noise and, consistently with typical experiments, to a small constant driving force \(\boldsymbol{f}(t) = \boldsymbol f_0 = f_0 \hat{\boldsymbol{x}} \) along the $x$ axis. The parameter values are the same as used in Section \ref{nongaussian}: $v_a=10^{-5}$ m/s (narrower curve) and $v_a=1$ m/s (broader curve) and $\tau_a=1$ s, $T=300$ K, $m=10^{-18}$ kg, $\gamma=10^{-9}$ kg/s and $f_0=10^{-12}$ N. The distribution for velocities along the $y$ axis looks the same but is completely centralized at $v_y=0$. (b) Simple illustration of heat distributions obtained through numerical Fourier transform of Eq.~\eqref{eq:ZQ} for the purely passive case with different combinations of initial and final temperatures. To be able to use a simpler numerical scheme, we used $\gamma=m=k_B=\tau=1$ in arbitrary units, with $T_0$ and $T_{\rm f}$ as indicated in the graph.} 
    \label{fig:Distr}
\end{figure}

For the active case, it is not as straightforward to modify Maxwell-Boltzmann as we did for the passive externally-driven case since activity breaks thermodynamic equilibrium locally and stochastically. Instead, we calculate the distribution of velocities for the AOUP in the underdamped regime by noticing that, at the velocity level, Gaussianity is preserved and then computing mean and variance explicitly; see Appendix \ref{B}. We find
\begin{equation}
    P_a(\boldsymbol v_0) \sim \exp\left(-\frac{\beta m }{2}\left(\boldsymbol v_0-\boldsymbol f_0/\gamma\right)^2\left(1+\frac{\beta m v_a^2}{(\tau_\gamma/\tau_a+1)}\right)^{-1}\right).\label{steadyactive}
\end{equation}
Here \(\tau_\gamma = m / \gamma\) defines the timescale relevant in the underdamped regime, which, in the overdamped regime, is a small number compared to the other timescales. 
%of the process \textcolor{red}{small compared with the persistence timescale or with $\tau$ or other?}. 
%We derive this velocity distribution in Appendix \ref{B}. 
Therefore, for the externally-driven AOUP, the stationary velocity distribution is still Gaussian, for each spatial dimension; see Fig.~\ref{fig:Distr}(a). By comparing against numerical results in \cite{shankar2018hidden} for other parameters, we find a perfect agreement (data not shown). Such Gaussianity of the velocity distribution is what allows us to find an analytical characteristic function for the heat.

\subsection{Fluctuations of kinetic energy}

The change in kinetic energy during the process has its own distribution due to its dependence on the initial and final velocities, which in turn are random variables as well. For both the passive and active cases, the velocity distributions are still Gaussian---see Eq.~\eqref{steadyactive}---with mean proportional to the driving and a variance denoted by $\sigma^2_{v_0}\equiv\langle \boldsymbol v_0^2 \rangle$, the expression for which depends on the case.

To further generalize our results, we also assume a non-isothermal process, which is common in stochastic heat engines with passive Brownian particles \cite{pires2023optimal, martinez2016brownian, blickle2012realization, de2019phase}. In this process, the final velocities equilibrate at a different temperature. Thus, the generalized joint distribution of the initial and final velocities reads
\begin{equation}
    P(\boldsymbol v_0,\boldsymbol v_\tau) \sim \exp\left(-\frac{(\boldsymbol v_0-\boldsymbol f_0/\gamma)^2}{2\sigma_{v_0}^2}-\frac{(\boldsymbol v_\tau-\boldsymbol f_0/\gamma)^2}{2\sigma_{v_\tau}^2}\right).
\end{equation}
For example, for a passive particle, we have \(\sigma_{v_0}^2 = 1/\beta_0 m\) and \(\sigma_{v_\tau}^2 = 1/\beta_\tau m\). 

From this joint velocity distribution, we proceed by calculating the Fourier transform of the distribution of changes in kinetic energy, that is, the characteristic function $Z_{\Delta K}(\lambda)$ \cite{paraguassu2022effects}, which reads
\begin{equation}
    Z_{\Delta K}(\lambda) =  \int d\boldsymbol v_0 \cdot d\boldsymbol v_\tau P(\boldsymbol v_0,\boldsymbol v_\tau)e^{-i\lambda \Delta K}.
\end{equation}
We can solve this integral for any spatial dimension $d$, since the velocities in a given direction are independent of those in the other directions. The characteristic function for $d$ dimensions is obtained by just multiplying the one-dimensional expression. The result is
\begin{eqnarray}
    Z_{\Delta K}(\lambda) = \exp\left({-\frac{ \boldsymbol{f}_0^2 m^2 \lambda^2 (\sigma_{v_0}^2 + \sigma_{v_\tau}^2)}{2 \gamma^2 (i + m \lambda \sigma_{v_0}^2) (-i + m \lambda \sigma_{v_\tau}^2)}}\right) \nonumber \\ \times\left[\left(1-i\lambda m\sigma_{v_0}^2\right)\left(1+i\lambda m\sigma_{v_\tau}^2\right)\right]^{-d/2}.\label{CFkineticenergy}
\end{eqnarray}
To find the distribution of changes in kinetic energy, we need to take the Fourier transform of the characteristic function. However, due to the dependence on $\lambda$ inside the exponential, an analytical solution is not possible. Nevertheless, for $d=2$, besides the aforementioned physical relevance, the square root outside the exponential disappears, simplifying the derivation of the statistical moments.

The characteristic function for the kinetic energy is
\begin{equation}
     Z_{\Delta K}(\lambda) = \frac{e^{-\frac{ f_0^2 m^2 \lambda^2 (\sigma_{v_0}^2 + \sigma_{v_\tau}^2)}{2 \gamma^2 (i + m \lambda \sigma_{v_0}^2) (-i + m \lambda \sigma_{v_\tau}^2)}} }{\left(1-i\lambda m\sigma_{v_0}^2\right)\left(1+i\lambda m\sigma_{v_\tau}^2\right)}.\label{CFkineticenergy2}
\end{equation}
where we assumed \(\boldsymbol{f}(t) = \boldsymbol f_0 = f_0 \hat{\boldsymbol{x}} \).
This characteristic function gives us all the moments of the kinetic energy change, and will be necessary to investigate the fluctuations of the heat.

\subsection{Fluctuations of work }\label{workdist}

In the passive case, the other contribution to the heat is the work; in the active case, it is the work together with an active heat contribution. Both quantities are Gaussian since they depend linearly on Gaussian quantities, \(\boldsymbol{v}(t)\) and \(\boldsymbol{f}_a(t)\). Since this contribution to heat is a Gaussian variable, we only need to have the mean and the variance. The characteristic function of the work is thus
\begin{equation}
    Z_W(\lambda) = \exp\left(-i\lambda\mu_W -\frac{\lambda^2\sigma_W^2}{2}\right).\label{CFwork}
\end{equation}
For the passive case, the mean can be easily calculated from Eqs.~\eqref{langevinp} and \eqref{workdef} by computing \(\langle \boldsymbol{v}(t) \rangle\), leading to
\begin{equation}
    \mu_W \equiv \langle W \rangle  = - \frac{1}{\gamma}\int_0^\tau \vert \boldsymbol{f}(t)\vert^2 dt,
\end{equation}
while the variance for the non-isothermal case is
\begin{equation}
    \sigma_W^2 = \frac{2}{\gamma}\int_0^\tau k_BT(t) \vert \boldsymbol{f}(t)\vert^2 dt .
\end{equation}
where the time-dependent temperature is due to the generalization to a non-isothermal process. In fact, to address a non-isothermal scenario, one just needs to allow for a time-dependent $T=T(t)$ in the passive noise correlation, Eq.~\eqref{passivenoise}. As an example, we choose a Heaviside function such that the temperature switches from \(T_0\) to \(T_\tau\) in the middle of a process of duration $\tau$, as previously investigated in the literature \cite{paraguassu2023heat}.

In the active case, the other contribution to the heat in Eq.~\eqref{activeheat} is a combination of work and an active heat contribution. Such combination gives a Gaussian variable, which we denote \(W_{\rm s}\). One can think of \(W_{\rm s}\) as shifted work in the sense that the active part is removed from the original work. We now show that \(W_{\rm s}\) in the active case obeys the same statistics as the work $W$ in the passive case.  The shifted work \(W_{\rm s}\) corresponds to the second term of the heat in Eq.~\eqref{activeheat}:
\begin{equation}
    W_{\rm s}[\boldsymbol{v}] = - \int_0^\tau \boldsymbol{f}(t) \cdot \left(\boldsymbol{v}(t) - \boldsymbol{f}_a \gamma^{-1}\right) dt.
\end{equation}
We can define a shifted velocity, \(\boldsymbol{V}(t) = \boldsymbol{v}(t) - \boldsymbol{f}_a \gamma^{-1}\). The stochastic nature of \(W_{\rm s}\) comes from this quantity. Now, we notice that the equation of motion for the active overdamped case can be rewritten in terms of \(\boldsymbol{V}(t)\) as
\begin{equation}
    \gamma \boldsymbol{V}(t) = \boldsymbol{f}(t) + \boldsymbol{\eta}(t),\label{shifteq}
\end{equation}
and the functional \(W_{\rm s}\) as
\begin{equation}
    W_{\rm s}[\boldsymbol V] = - \int_0^\tau \boldsymbol{f}(t) \cdot \boldsymbol{V}(t) dt.
\end{equation}
Due to Eq.~\eqref{shifteq}, the statistics of \(\boldsymbol{V}(t)\) are identical to the statistics of $\boldsymbol v(t)$ in the passive motion equation, Eq.~\eqref{langevinp}. Consequently, the statistics of \(W_{\rm s}\) will be the same as those of the work in the passive case. That is, \(\langle W \rangle = \langle W_{\rm s} \rangle\) and \(\sigma_W^2 = \sigma_{W_{\rm s}}^2\). 

Therefore, the effect of the active noise appears only in the kinetic energy, particularly in the variances of the initial and final velocities. This means that without the kinetic energy, the heat distribution would wrongly be the same in both the passive and active cases. 

\subsection{Heat fluctuations}

The heat depends linearly on the kinetic energy and the work (for the active case, it is the shifted work $W_{\rm s}$): $Q=\Delta K + W$. The work and the kinetic energy change are uncorrelated variables in the overdamped regime, where the kinetic energy is wrongly ignored \cite{paraguassu2022effects}. To see this, notice that the work depends on the particle's trajectory, but not on the initial and final \textit{velocities} $\boldsymbol v_0$ and $\boldsymbol v_\tau$, that is, we can rewrite the work Eq.~\eqref{workdef} in terms of the initial and final \textit{positions} $\boldsymbol x_0$ and $\boldsymbol x_\tau$
% $W= - f_0(x_\tau-x_0) + \int_0^\tau \dot f x(t) dt$, 
while the kinetic energy depends only on the initial and final velocities. Now, in general, these boundary conditions $\boldsymbol x_0, \boldsymbol x_\tau, \boldsymbol v_0$, and $\boldsymbol v_\tau$ will be correlated, as should appear in their joint distribution $P(\boldsymbol x_\tau, \boldsymbol x_0, \boldsymbol v_\tau, \boldsymbol v_0)$. However, in the overdamped limit, where $\tau_\gamma$ is much smaller than the time of the process $\tau$, we achieve $P(\boldsymbol x_\tau, \boldsymbol x_0, \boldsymbol v_\tau, \boldsymbol v_0) = P(\boldsymbol x_\tau,\boldsymbol x_0)P(\boldsymbol v_\tau,\boldsymbol v_0)$. 

As a consequence of the statistical independence of $\Delta K$ and $W$, their joint distribution becomes
\begin{equation}
    P_{\Delta K, W}(\Delta K, W) = P_W(W)P_{\Delta K}(\Delta K).
\end{equation}
Since by construction $Q=\Delta K + W$, the heat distribution obeys
 \begin{align}
    P_Q(Q) &= \int\!\!\!\int \! d\Delta K dW\; \delta(Q-(\Delta K+W)) P_{\Delta K, W}(\Delta K, W) \nonumber \\
           &= \int d\Delta K P_{\Delta K}(\Delta K) P_W(Q-\Delta K). \label{heateq}
\end{align}
However, as previously indicated, \(P_{\Delta K}(\Delta K)\) is not analytically accessible to us. Nevertheless, all heat fluctuations are also encoded in the characteristic function of the heat. Due to the statistical independence of work and kinetic energy, the characteristic function will be given by
\begin{equation}
    Z_Q(\lambda) = Z_{\Delta K}(\lambda) Z_W(\lambda),
    \label{eq:ZQ}
\end{equation}
where $Z_{\Delta K}$ and $Z_W$ are given in Eqs.~\eqref{CFkineticenergy} and \eqref{CFwork}, respectively. 

The heat distribution can then be obtained by calculating the Fourier transform of $Z_Q(\lambda)$. However, due to the polynomial dependence of $\lambda$ inside the exponential in $Z_{\Delta K},$ an analytical solution remains not possible. We thus proceed numerically and merely illustrate the emerging heat distribution in Fig.~\ref{fig:Distr}(b). To be able to use a simpler numerical scheme, we chose arbitrary units in which the parameter values are close to unity. One can see that the distribution is not Gaussian, as for $T_0\neq T_f$ we have clearly an asymmetry. For different temperatures (non-isothermal case), the heat distribution becomes asymmetric. One way to see that is by comparing both sides of the non-isothermal distributions to the isothermal one in Fig.~\ref{fig:Distr}(b). Also, such asymmetry in the heat distribution can be measured by its non-zero skewness, which in turn implies non-Gaussian behavior. As demonstrated in Appendix \ref{C}, the skewness depends solely on the difference in velocity variances between the initial and final time instants. As we shall see, this distribution features unique mean, variance, skewness, and excess kurtosis which offer quantitative metrics of its deviation from a Gaussian distribution.

Another quantity of interest is the joint characteristic function of the work and heat. Since what breaks the Gaussianity is the kinetic energy, the work can be seen as the heat if the kinetic energy were to be ignored. Therefore, knowing the statistical correlation between the two stochastic variables informs us about the effects of kinetic energy on heat fluctuations.

The way to compute the correlation between heat and work analytically is through the joint characteristic function of both quantities, $Z_{Q,W}(\lambda, \lambda')$. This characteristic function can be derived by noticing that the joint distribution of heat and work obeys
%having the joint distribution of heat and work, $P_{Q,W}(Q,W)$, which is already given inside the integral of Eq.~\eqref{heateq}; using the Dirac delta to remove the integral of the kinetic energy, the left part of the integrand will be the joint distribution
\begin{equation}
    P_{Q,W}(Q,W) = P_{\Delta K}(Q-W) P_W(W).
\end{equation}
With that, and knowing the characteristic functions of the work and kinetic energy, we can find the joint characteristic function (see Appendix \ref{jointApp}):
\begin{equation}
    Z_{Q,W}(\lambda,\lambda') = Z_{\Delta K}(\lambda)Z_W(\lambda +\lambda') 
\end{equation}
To find any joint moment of the work and heat, we only need to have $Z_{Q,W}(\lambda, \lambda')$ and use that
\begin{equation}
    \langle Q^n W^m\rangle = i^{n+m} \frac{\partial^{n+m} Z_{Q,W}(\lambda, \lambda') }{\partial^n \lambda \partial^m \lambda'} \bigg|_{\lambda, \lambda' \rightarrow 0}. \label{eq:joint-distri}
\end{equation}

\section{non-Gaussian Heat fluctuations}\label{nongaussian}

We have observed that the heat distribution is not Gaussian, due to the influence of kinetic energy. But to what extent is the effect of kinetic energy relevant? Although the heat distribution obtained in the previous section is not strictly Gaussian, it does contain some degree of Gaussianity, particularly for the illustration parameters used. To analyze how much we deviate from Gaussian behavior, and consequently how relevant the kinetic energy is, we will investigate the excess kurtosis and the Pearson correlation coefficient for passive and active systems.

\subsection{Passive systems}

We consider an adapted experimental situation involving an overdamped silica particle \cite{pires2023optimal}. Specifically, we choose a particle of mass $m = 10^{-18}$ kg subject to damping coefficient $\gamma = 10^{-9}$ kg/s, and a typical process interval on the order of milliseconds, with $\tau = 10^{-3}$ s.

In a passive scenario, a temperature protocol is often used as one is frequently interested in non-isothermal processes \cite{pires2023optimal}. For simplicity, we assume a Heaviside temperature protocol \(T(t) = T_0 + T_{\rm f} \Theta(\tau/2)\), where $\Theta(t)$ is the Heaviside function. This particular temperature protocol was used, for instance, in \cite{pires2023optimal, paraguassu2023heat2}. For the external driving protocol, \(\boldsymbol{f}(t)\), we assume a constant force, \(\boldsymbol{f}(t) = f_0 \hat{\boldsymbol{x}} \). This force can be applied using electrodes \cite{kremer2024all, kremer2024perturbative}, and it is typically on the order of piconewtons, so we choose \(f_0 = 10^{-15}\) N.

\subsubsection{Excess kurtosis: Passive case}

The excess kurtosis measures the frequency and intensity of rare events in a distribution, reflecting the behavior of the tails of the distribution. In other words, it measures the ``tailedness'' of a distribution, indicating whether the data are heavy-tailed (positive kurtosis) or light-tailed (negative kurtosis) compared to a Gaussian distribution. Therefore,
for a Gaussian distribution, the excess kurtosis is defined as zero. Any value different from zero indicates a deviation from Gaussian behavior. Thus, we can use the excess kurtosis to assess the importance of the kinetic energy since it breaks Gaussianity. 

We can calculate the excess kurtosis by integrating the distribution directly or by using its Fourier transform, i.e., the characteristic function. We choose the latter; in Appendix \ref{C} we detail the derivation of central moments via the characteristic function. The result for the excess kurtosis is
\begin{equation}
     \kappa = \frac{6 \gamma^2 m^4 \left(\gamma^2 \left(\sigma_{v_0}^8 + \sigma_{v_\tau}^8\right) + 4 f_0^2 \left(\sigma_{v_0}^6 + \sigma_{v_\tau}^6\right)\right)}{\left(2 f_0^2 m^2 \left(\sigma_{v_0}^2 + \sigma_{v_\tau}^2\right) + \gamma^2 \left(m^2 \left(\sigma_{v_0}^4 + \sigma_{v_\tau}^4\right) + \sigma_w^2\right)\right)^2}.
     \label{excesskurtosis}
\end{equation}
In this form, this expression for the excess kurtosis is valid for both the passive and active cases, with distinctions appearing in the variances of the velocities. First, we analyze the excess kurtosis for $T_0=T_{\rm f}$ to see the effect of the driving on (the tail of) the distribution. In Fig.~\ref{fig:PassiveKappaRho}(a), we plot excess kurtosis versus driving force in the range $f_0\in[10^{-15},\,10^{-12}]$ N. The minimum value $\sim 10^{-15}$ N is typical for electric forces produced by electrodes in experiments with optical tweezers  \cite{kremer2024all}. Moreover, one can find Brownian particles subject to external driving forces with values across various orders of magnitude  depending on the context and origin of the force, e.g., gravitational, electrostatic, hydrodynamic, van der Waals, and cellular forces.

For $f_0 \to 0$, the kinetic energy is the only contribution to the heat and therefore the excess kurtosis plateaus at $\kappa=3$. An excess kurtosis of $3$ is equivalent to a kurtosis of $6$, which is exactly double the kurtosis of a Gaussian distribution. This is because the kinetic energy depends on the square of a Gaussian variable, so when the heat distribution is dominated by kinetic energy effects, the kurtosis doubles with respect to the Gaussian case. An excess kurtosis greater than zero corresponds to leptokurtic distribution \cite{paraguassu2023heat2}, that is, a distribution that has more rare events than the Gaussian case, which in turn is called mesokurtic distribution. Upon increasing $f_0$, the contribution from kinetic energy effects to the heat becomes progressively less relevant, leading to a smooth crossover transition above which we reach a plateau with $\kappa=0$ and the heat distribution becomes Gaussian. For higher temperatures, kinetic energy effects are more pronounced and therefore the transition driving force needed to overcome such effects becomes larger.

In Figs.~\ref{fig:PassiveKappaRho}(b) and (c), we show the excess kurtosis versus the initial and final temperatures of the stochastic trajectory for two values of the force and the temperature range used in \cite{martinez2016brownian}. We see that as we increase the temperature difference, the excess kurtosis increases, meaning that the kinetic energy becomes more relevant in the statistics of the heat. This is expected since, as previously hinted, for non-isothermal processes the heat distribution becomes asymmetric, further departing from Gaussian behavior.

% Apart from the asymmetry, it is worth noticing that the excess kurtosis has a threshold value of $\kappa = 3$ for isothermal processes. Also, interestingly, the force has a minimum value which affects the excess kurtosis. This makes sense since for $f_0 \to 0$ the kinetic energy becomes the only contribution to the heat. Note that, an excess kurtosis of 3 is equivalent to the standard kurtosis of 6, which is exactly double the kurtosis of a Gaussian distribution. This is reasonable since the kinetic energy depends on the square of a Gaussian variable, so its kurtosis becomes the double of the Gaussian case.    

\subsubsection{Pearson correlation coefficient: Passive case}

A different way of assessing the relevance of the kinetic energy is through the Pearson correlation coefficient. This quantity measures the correlation between two stochastic variables and is equal to $1$ for perfect correlation. The Pearson correlation coefficient is defined as
\begin{align}
    \rho = \frac{\langle W Q \rangle - \langle W \rangle \langle Q \rangle}{\sigma_W \sigma_Q}.\label{pearson}
\end{align}
In particular, a value of $1$ would mean that work and heat are perfectly correlated and therefore imply that the kinetic energy change is irrelevant. Conversely, a small Pearson correlation coefficient means that the work does not play an important role in the heat compared to the kinetic energy change. The correlation $\langle W Q \rangle$ can be obtained using Eq.~(\ref{eq:joint-distri}). The resulting Pearson correlation coefficient is just the ratio between the variance of the Gaussian case (the work variance) and the variance of the actual heat distribution:
\begin{equation}
    \rho = \frac{\sigma_W}{\sigma_Q}. \label{eq:Pearson}
\end{equation}
If $\sigma_Q\approx\sigma_W$, one can ignore the kinetic energy contribution since $\rho \approx 1$. 

Using $\sigma_Q^2 = \sigma_{\Delta K}^2+\sigma_W^2$ (see Appendix \ref{C}), we write
% \begin{equation}
%     \rho = \left[1+\left(\frac{\sigma_{\Delta K}}{\sigma_W}\right)^2\right]^{-1/2}. \label{eq:Pearson-rewritten}
% \end{equation}
\begin{equation}
    \rho = \frac{1}{\sqrt{1+\left(\sigma_{\Delta K}/\sigma_W\right)^2}}. \label{eq:Pearson-rewritten}
\end{equation}
As we increase the fluctuations in the change of kinetic energy, the correlation decreases, while by increasing the fluctuations of the work, the correlation increases, leading to less relevant kinetic energy effects.

% For the values discussed here, we find that the variance of the change in kinetic energy is on the order of $\sigma_{\Delta K}^2  \sim (k_B T)^2$ (see Eq.~(\ref{eq:sigmaK}) and Eq.~(\ref{eq:variancia})), while the work variance is
% $\sigma^2_W \sim f_0^2\tau k_BT/\gamma  = 10^{-24}$ J $\times$ $k_B T$. At room temperature, the ratio between the variances in the passive case is
% \begin{equation}
%     \left(\frac{\sigma_{\Delta K}}{\sigma_W}\right)^2 \sim 10^3 \gg 1.
% \end{equation}
%  This corresponds to a small Pearson correlation coefficient of $\rho \sim 10^{-2}$, meaning that the correlation between work and heat is low. Therefore, fluctuations of kinetic energy change dominate over those of the work. 

 \begin{figure*}[htbp]
    \centering
    \includegraphics[width=0.95\textwidth]{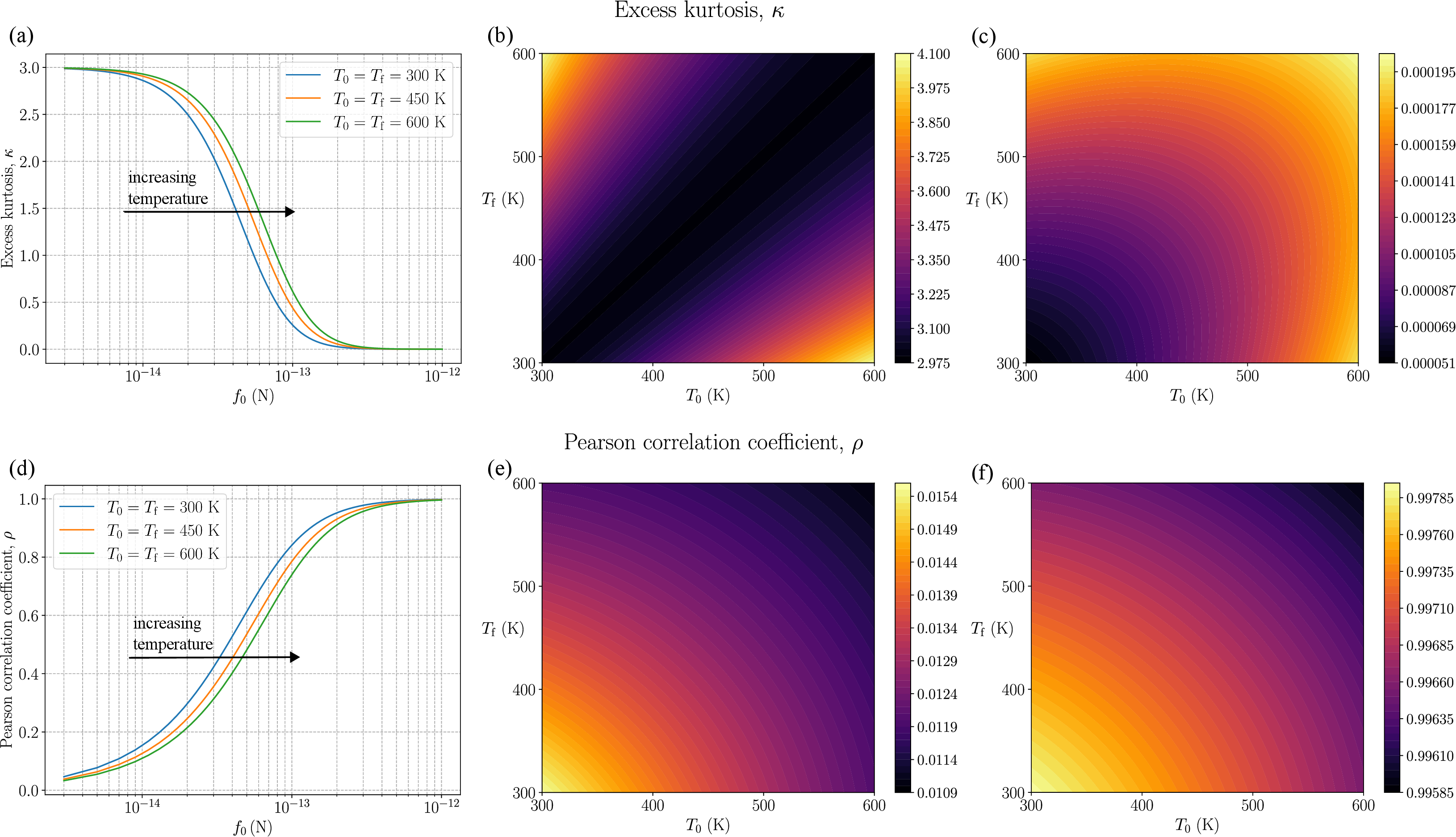}
    \caption{Excess kurtosis [Eq.~\eqref{excesskurtosis}] and Pearson correlation coefficient [Eq.~\eqref{eq:Pearson-rewritten}] for the purely passive case. (a) Excess kurtosis versus $f_0$ for $T_0=T_{\rm f}=300$ K, $450$ K, and $600$ K. Excess kurtosis versus $T_0$ and $T_{\rm f}$ for (b) $f_0=10^{-15}$ N and (c) $f_0=10^{-12}$ N. (d) Pearson correlation versus $f_0$ for $T_0=T_{\rm f}=300$ K, $450$ K, and $600$ K. Pearson correlation versus $T_0$ and $T_{\rm f}$ for (e) $f_0=10^{-15}$ N and (f) $f_0=10^{-12}$ N. Other parameters values: $m=10^{-18}$ kg, $\gamma=10^{-9}$ kg/s, $\tau=10^{-3}$ s and $k_B=1.38 \times 10^{-23}$ J/K. In (a) and (d), temperature increases as indicated by the arrows.}
    \label{fig:PassiveKappaRho}
\end{figure*}

Generally, both the Pearson correlation coefficient and the excess kurtosis reveal the same crossover transition behavior. When the external driving force increases or temperature decreases, there is a transition from a regime dominated by kinetic energy effects and non-Gaussianity of the heat distribution to a regime dominated by external work with a Gaussian heat distribution. However, these two metrics still provide different information. This can be seen by noticing their behavior in Figs.~\ref{fig:PassiveKappaRho}(b) and (d). For the isothermal behavior ($T_0 = T_{\rm f}$ diagonal), the Pearson correlation coefficient is capable of capturing a more significant change than the excess kurtosis as temperature is varied within the same temperature range. Presumably, this is because, while the kurtosis measures tailedness, the Pearson correlation coefficient is concerned with a lower statistical moment, i.e., the variances. In fact, variances are expected to be more quickly affected as we vary, say, temperature than the distribution tails. A similar result and observation can be made for the active case below.

% The correlation provides a different information about the non Gaussianity. Differently from the excess kurtosis, it deals with the variance of the distributions. Here, the Pearson coefficient lead to the same conclusions that we got from the excess kurtosis, but informing a different behavior. 

% In Fig.~\ref{fig:PassiveKappaRho}(d), we show the Pearson correlation coefficient for different values of temperature over the driving $f_0$. As the driving increases we achieve a perfect correlation $+1$. This means that the heat is perfect correlated with the work, and therefore can be described by a Gaussian distribution. Thus the kinetic energy is irrelevant as the force increases. 

% For a non-isothermal process, $\Delta T \neq 0$, the difference in temperature decrease the Pearson coefficient. Increasing the distance between the two variables. This behavior is shown in Fig.~\ref{fig:PassiveKappaRho}(e) and Fig.~\ref{fig:PassiveKappaRho}(f) this behavior. 

\subsection{Active systems}

For the active case, we restrict our analysis to the most common isothermal scenario.

\subsubsection{Excess kurtosis: Active case}

Activity introduces a velocity scale $v_a$ that controls the intensity of the active noise, as discussed in Section \ref{modelandthermo}. The more general formula for the excess kurtosis, Eq.~\eqref{excesskurtosis}, remains the same, but now with $\sigma_{v_0}^2$, or $\sigma_{v_\tau}^2$, as is given in the fully overdamped version of Eq.~(\ref{eq:variancia}):
\begin{equation}
    \sigma^2_v = \frac{1}{m\beta}\left( 1+ \frac{\beta mv_a^2}{2}\right). \label{eq:variancia}
\end{equation}
Interestingly, $\tau_\gamma/\tau_a$ is close to zero and therefore $\tau_a$ disappears at this level in the overdamped limit. (In Fig.~\ref{fig:Distr}(a) we used a typical value in the range $\tau_a=0.1$---$1$ s \cite{maggi2014generalized} as indicated therein.)
Notice that we can define an active kinetic energy scale as \(K_a = \frac{mv_a^2}{2}\). The active velocity therefore increases the fluctuations of the kinetic energy. Since the shifted work has the same statistics of the work in the passive case, $v_a$ only contributes to the kinetic energy. This leads to a deviation of the heat distribution from the Gaussian behavior caused by activity.

\begin{figure}[htbp]
    \centering
    \includegraphics[width=0.95\columnwidth]{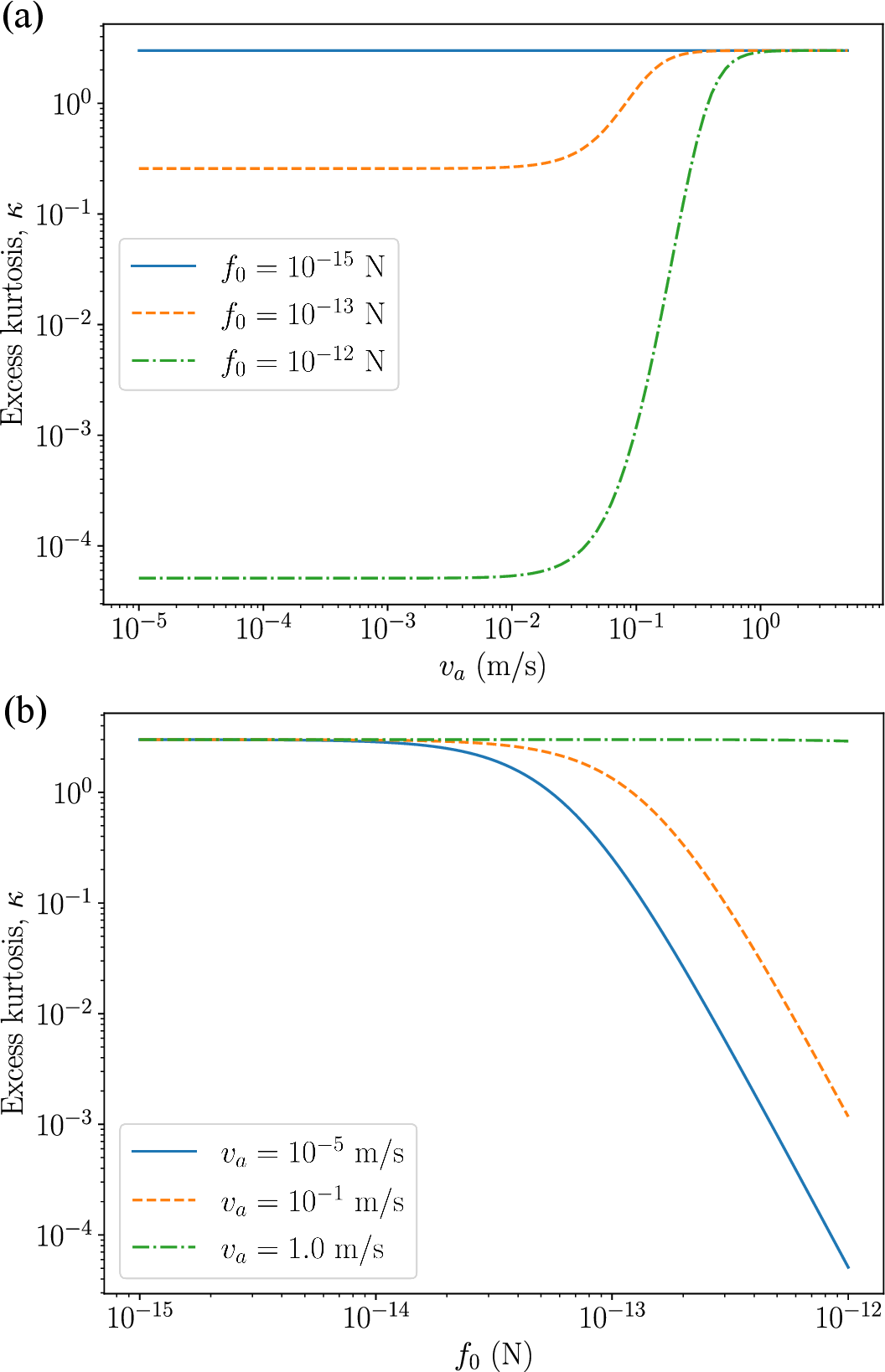}
    \caption{Excess kurtosis [Eq.~\eqref{excesskurtosis}] for the active case (a) versus $v_a$ for $f_0=10^{-15}$ N (solid line), $f_0=10^{-13}$ N (dashed line), and $f_0=10^{-12}$ N (dot-dashed line) and (b) versus $f_0$ for $v_a=10^{-5}$ m/s (solid line), $v_a=10^{-1}$ m/s (dashed line), and $v_a=1$ m/s (dot-dashed line). Other parameters values: $m=10^{-18}$ kg, $\gamma=10^{-9}$ kg/s, $\tau=10^{-3}$ s, $T_0=T_{\rm f}=300$ K, and $k_B=1.38 \times 10^{-23}$ J/K.}
    \label{fig:KappaActive}
\end{figure}

In Fig.~\ref{fig:KappaActive}(a), we show the excess kurtosis versus the active velocity $v_a$ for three different values of the force $f_0$. The range of $v_a$ starts from the order of magnitude used in the particular microparticle experiment of Ref.~\cite{maggi2014generalized}, $v_a=10^{-5}$ m/s, and moves across orders of magnitude as can be accessed in other active matter systems \cite{bechinger2016active}. For sufficiently low driving force, the excess kurtosis remains saturated at $\kappa=3$, in such a way that $v_a$ does not affect the excess kurtosis $\kappa$. In this case, the system is already at its maximum excess kurtosis, in a regime dominated by kinetic energy effects. On the other hand, at high driving force, as we increase $v_a$, initially the excess kurtosis remains almost zero (corresponding to a Gaussian heat distribution, with no kinetic energy effects) but then undergoes a crossover transition to the regime where kinetic energy effects are relevant and the heat distribution is non-Gaussian again.
%Is worth pointing out that in figure y a) we extrapolate the range of $v_a$. Typical values include $v_a=10^{-5}$ m/s and $\tau_a=0.1$---$1$ s \cite{maggi2014generalized}. We also include larger values of $v_a$ in order to show the crossover from a regime dominated by work fluctuations to one dominated by fluctuations in kinetic energy change. This indicates that overdamped active particles at larger active velocity scales could have heat fluctuations dominated by kinetic energy changes through activity effects.

Similar conclusions could be drawn by looking at the plots in Fig.~\ref{fig:KappaActive}(b) where we choose three different values of $v_a$, and vary the driving force $f_0$. 
%As $f_0$ increases, the fluctuations of work become more important and thus assymptotically 
%\begin{align}
%    \lim_{f_0 \to \infty}\kappa = 0. 
%\end{align}
%This means that in the work dominated regime, we recover the Gaussian behavior for the excess kurtosis. 
Analyzing both plots, we observe a competition between the activity $v_a$, promoting kinetic energy effects, and the driving $f_0$, promoting external work effects. This can be summarized by noticing that
%As already mentioned, by increasing $f_0$ we privilege work, and on the other side, increasing $v_a$ lead us to the case where the excess kurtosis is associated to the double of the Gaussian case. Meaning that the kinetic energy is dominating the statistics of the distribution. 
\begin{align}
    v_a \to \infty &\Rightarrow P\left( Q \right) = P_{\Delta K}\left( Q \right) \\
    f_0 \to \infty &\Rightarrow P\left( Q \right) = P_{W}\left( Q \right)
\end{align}
For intermediate values of $v_a$ and $f_0$, both thermodynamical quantities, $W$ and $\Delta K$, affect the heat statistics. In case one wants to average out kinetic energy changes, one would have
\begin{align}
    P\left(Q\right) = \langle P_{W}\left( Q- \Delta K \right) \rangle_{\Delta K},
\end{align}
where $\langle \dots \rangle_{\Delta K}$ means that we are taking the average over solely the kinetic energy distribution.

\subsubsection{Pearson correlation coefficient: Active case}

For the active case, we have the same expression for the Pearson correlation coefficient, Eq.~(\ref{eq:Pearson}). The variances of the work or the shifted work are equal while, as discussed above, the variance of the kinetic energy change is shifted by the active kinetic energy [see Eqs.~\eqref{eq:sigmaK} and \eqref{eq:variancia}], that is,
% \begin{equation}
%     \sigma_{\Delta K}^2 = \left(k_B T\right)^2 \left(1 + \frac{K_a}{k_B T}\right)\left( 1+\frac{K_a}{k_B T} + \frac{m (f_0/\gamma)^2}{k_B T} \right),
% \end{equation}
\begin{equation}
    \sigma_{\Delta K}^2 = 2\left(k_B T + K_a\right)\left( k_B T+K_a + m f_0^2 /\gamma^2 \right),
\end{equation}
which is valid for $\tau_a \gg \tau_\gamma$. 
%With the experimental parameter values used here, we have \(K_a \sim 10^{-28}\) J and \(k_B T \sim 10^{-21}\) J. Thus, for these parameters, the active noise affects the fluctuations in kinetic energy change only slightly. Qualitatively, it increases the fluctuations in kinetic energy change. In fact, the kinetic energy remains the dominant fluctuation in the heat. 

\begin{figure}[htbp]
    \centering
    \includegraphics[width=0.95\columnwidth]{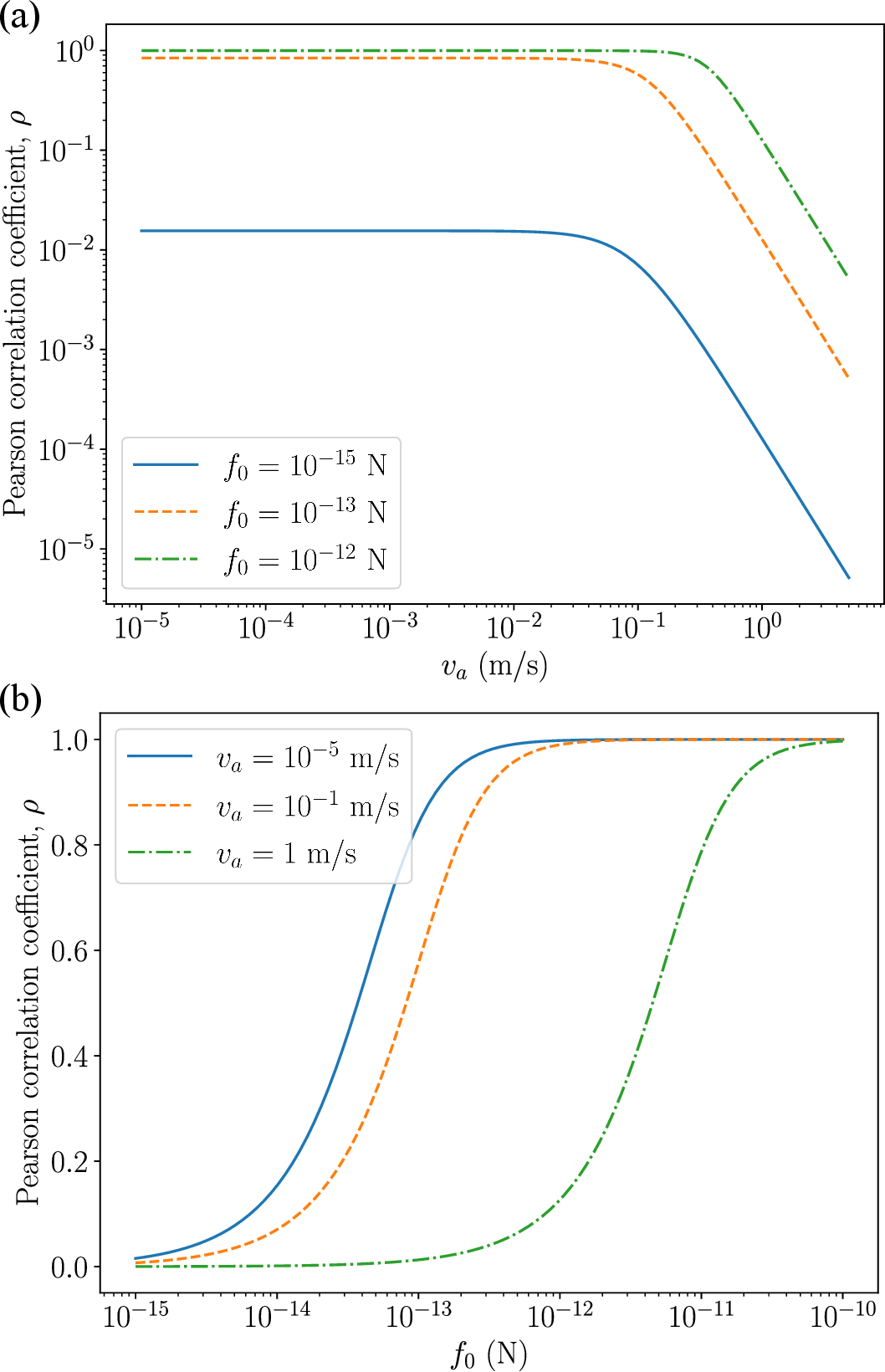}
    \caption{Pearson correlation coefficient [Eq.~\eqref{eq:Pearson-rewritten}] for the active case (a) versus $v_a$ for $f_0=10^{-15}$ N (solid line), $f_0=10^{-13}$ N (dashed line), and $f_0=10^{-12}$ N (dot-dashed line) and (b) versus $f_0$ for $v_a=10^{-5}$ m/s (solid line), $v_a=10^{-1}$ m/s (dashed line), and $v_a=1.0$ m/s (dot-dashed line). Other parameters values: $m=10^{-18}$ kg, $\gamma=10^{-9}$ kg/s, $\tau=10^{-3}$ s, $T_0=T_{\rm f}=300$ K, and $k_B=1.38 \times 10^{-23}$ J/K.}
    \label{fig:RhoActive.eps}
\end{figure}

%If we allow the active velocity to vary arbitrarily, we can see this effect on the statistics of the heat. In figure z a) we have the Pearson coefficient varying $v_a$ for different values of force. The maximum value of 1 is obtained for $f_0 = 10^{-12}$ N. 
Figs.~\ref{fig:RhoActive.eps}(a) and (b) show the Pearson correlation coefficient as a function of $v_a$ and $f_0$, respectively. Asymptotically, for high activity, the heat becomes uncorrelated with the work: $\rho \to 0$ as $v_a \to \infty$. For small driving force and $v_a$, $\rho$ is small but not zero, in a sort of crossover regime. In this scenario, the heat is already not Gaussian distributed. As we increase the activity $v_a$, the Pearson correlation coefficient goes to zero. Notice, however, that, since activity is an additional source of noise, the limit $v_a \to 0$ does not lead to a perfect correlation between work and heat as the passive noise is still present. Therefore, with activity, the regime transitions are no longer between extreme scenarios for the excess kurtosis and Pearson correlation coefficient.

As in the passive case, we highlight that, although the regime transition is captured by both excess kurtosis and Pearson correlation coefficient, these metrics deliver different information; upon varying $v_a$, Fig.~\ref{fig:KappaActive}(a) and Fig.~\ref{fig:RhoActive.eps}(a) show no transition for the excess kurtosis but do show a transition for the Pearson correlation coefficient. Again, this is presumably because the Pearson correlation coefficient feels changes of parameters related to the heat distribution variance more sensitively (or ``first'') than the excess kurtosis does since the latter is related to the tails of the distribution. For instance, compare Fig.~\ref{fig:KappaActive}(b) and Fig.~\ref{fig:RhoActive.eps}(b) for the variation against $f_0$.
%In this regime, the heat statistics is dominated by the kinetic energy. There is an intermediate regime, where both $v_a$ and $f_0$ affect the Pearson coefficient, in which the driving and the activity compete. For instance, one could use the external driving force $f_0$ and artificial generated noise \cite{danger2000efficient} $v_a$ as tuning parameters to realize a situation in which the Pearson coefficient acquire values smaller than one. 

%\section{Influence of activity type}

\section{Regime transition parameters}
Our initial hypothesis is that, for example, in the passive isothermal case, the driving force and the thermal noise compete to take the system into a particular stochastic thermodynamics regime. In principle, one could therefore predict where the transition will occur by simply equating the driving and stochastic forces that appear in our equation of motion, Eq.~\eqref{langevinp}. More precisely, since we are dealing with stochastic equations, one should stochastically integrate in time (over $\tau$). By doing so, after dividing by $\tau$, we obtain an estimated transition at
\begin{equation}
f_0  = \sqrt{\frac{2 \gamma k_B T}{\tau}}
\end{equation}
Although this simple procedure gives an excellent estimate of the middle of the crossover transition (as we checked), it gives no information regarding how smooth is the transition. On the other hand, Figs.~\ref{fig:PassiveKappaRho}, \ref{fig:KappaActive}, and \ref{fig:RhoActive.eps} show that the transition can be quite smooth and thus occur over a range of the control parameter that spans orders of magnitude. Therefore, knowing the middle point of the crossover transition is insufficient. This shows why metrics such as the excess kurtosis or Pearson correlation coefficient are needed in order to actually be able to estimate how deep into a particular regime our overdamped particle lives. 

Nonetheless, the above calculation of the transition parameters can serve as a way to partially check the intricate procedure developed here to obtain at least one of our metrics, the Pearson correlation coefficient. In the passive isothermal case, we can observe that the middle of the transition occurs for $\rho=1/2$. Solving for this condition and using that $\tau\gg\tau_\gamma$ yields
\begin{equation}
f_0  = \sqrt{\frac{2 \gamma k_B T}{3\tau}}.
\end{equation}
Differing only by a factor of $\sqrt{3}$, these two approaches give the same order of magnitude for the transition. This corresponds to a very good agreement considering that the two approaches effectively involve different orders of statistical moments of the stochastic quantities. Analyses in the same spirit can be made for the active and non-isothermal cases.

\section{Concluding remarks}\label{discussion}
In this paper we described the statistical behavior of heat exchanged between a driven overdamped particle and its surrounding thermal bath for both passive and active particles. We focused on the effects of changes in kinetic energy which had been overlooked in the past under the assumption that no terms involving mass should be relevant in the overdamped limit. As we show, although mass is indeed absent from the equation of motion itself, it can still be relevant in the energetic dynamics. Using two quantities---the Pearson correlation coefficient and the excess kurtosis---we show that this is true even in the presence of driving forces that could reduce stochasticty and thus the relevance of the kinetic energy contribution. In particular, we demonstrate that the distribution of heat changes from a Gaussian to a non-Gaussian behavior, regimes where it is dominated by external work and kinetic energy, respectively. We also identify an intermediate regime where both quantities contribute to the exchanged heat statistics over decades of the control parameters and where the distribution is also non-Gaussian. 

While the excess kurtosis informs about rare events associated with the tail of a distribution, the Pearson correlation coefficient deals with covariances. We found that the excess kurtosis and the Pearson coefficient lead us to the same conclusion regarding the overal regime transition, but since they involve different staistical moments, one of them can be more sensitive than the other to changes in control parameters over a given range.
For passive isothermal systems, the excess kurtosis $\kappa$ decreases when one increase the driving force $f_0$ from a maximum value of $\kappa = 3$. For non-sothermal processes, changes in kinetic energy become more pronounced and the heat distribution is asymmetric. As a result, $\kappa>3$. 
%In this case we obtain a positive excess kurtosis that increases with temperature, showing a deviation from Gaussian behavior and indicating the leptokurtic \cite{paraguassu2022effects,paraguassu2023heat2,paraguassu2023heat} statistical nature of the heat and the importance of the kinetic energy in this process. In the case of the Pearson coefficient we realize that for non-isothermal process $\Delta T \neq 0$, by raising the temperature difference we decorrelate both variables, decreasing the Pearson coefficient, which indicates non-Gaussian behavior.

Investigating the role of the activity, we found that its effect is to qualitatively increase the fluctuations in kinetic energy change, therefore promoting non-Gaussianity of the heat distribution. On the other hand, the external driving generates a more Gaussian behavior. We showed that, therefore, there exists a crossover regime where driving and activity compete against each other. In this regime, the heat distribution remains non-Gaussian. 
%We notice that the fact that activity promotes non-Gaussianity (i.e., kinetic energy change effects) may be specific to AOUPs, where the velocity distribution in each direction is Gaussian \cite{shankar2018hidden}. This differs from the ABP model, where the velocity distribution is bimodal. For ABPs, one would expect that, for high persistence times compared to the heat measurement time $\tau$, activity could actually act similarly to an external driving force and therefore \textit{reduce} non-Gaussianity. 
A thorough investigation of the effect of changes in kinetic energy on the heat distribution for other models of active particles is interesting future avenue. This includes generalized run-and-tumble models for various types of bacteria \cite{villa2020run}. 
%\textcolor{red}{Pablo to Pablo: review story on type of activity and heat definition to answer whether activity promotes non-Gaussianity}

% Within our framework, our results show that the only way to be able to neglect the kinetic energy is by significantly increasing the scale of the driving force. In fact, it was only by increasing the driving force by two orders of magnitude above experimental values that we could finally observe the predominance of work over kinetic energy change in the heat statistics, as demonstrated through the excess kurtosis and the Pearson. Our working hypothesis is thus confirmed: a reduction in stochasticity at the spatial level translates into the energetic level and makes kinetic energy less important. With the parameters used above, for a driving force $f_0=10^{-15}$ N, and assuming that the thermal forces are of the order of $\sqrt{2\gamma k_B T}=10^{-14}$ N, one could anticipate that kinetic energy effects will dominate heat fluctuations, which is observed indeed. On the other hand, for $f_0=10^{-13}$ N, the opposite is expected, and this is also confirmed by our analyses of the heat statistics.

In summary, our results provide evidence that the effect of kinetic energy on the heat statistics for overdamped particles, where inertia does not play a role in the movement dynamics, cannot be readily ignored even in the presence of driving forces which reduce stochasticity. On the contrary, when calculating the heat fluctuations and using two metrics of non-Gaussianity and thus of the effects of kinetic energy, we found that there is a regime where kinetic energy effects could dominate the fluctuations over the driving force work, both for passive and active particles. Finally, we showed quantitatively that, although a simpler analysis can give a good estimate regarding where the transition between the two regimes occurs, only a more comprehensive analysis as the one presented here can capture how deep into a particular regime the system is. We hope that our results can lead to new analyses on the quantitatively relevant role of kinetic energy effects in overdamped stochastic systems.
%Moreover, we found out that, for the experimental values considered here, active noise has only a small influence on kinetic energy effects. Given that \(\sigma_{\Delta K} \sim k_B T\), what really dominates is the thermal energy of the passive thermal bath. 

%We have shown that, even when interacting with the system through a driving force, we cannot ignore kinetic energy in the heat expression for overdamped systems. Concluding that, using the Sekimoto expression naively in overdamped systems leads to incomplete results that cannot be ignored for real-life parameters.

\section*{Acknowledgments:}  This work is supported by the Brazilian agencies CAPES and CNPq. P.V.P acknowledges Diego Charpenel and the Funda\c{c}\~ao de Amparo \`a Pesquisa do Estado do Rio de Janeiro (FAPERJ Process SEI-260003/000174/2024). This study was financed in part by Coordena\c c\~ ao de Aperfei\c coamento de Pessoal de N\' ivel Superior - Brasil (CAPES) - Finance Code 001. Also, P.V.P. would like to thank ICTP-SAIFR (FAPESP grant 2021/14335-0) where part of this work was done. R.A.\ is supported by a Post-Doctoral Fellowship No.\ 2023/05765-7 granted by São Paulo Research Foundation (FAPESP), Brazil. P.d.C.\ was supported by Scholarships No.\ 2021/10139-2 and No.\ 2022/13872-5 and ICTP-SAIFR Grant No.\ 2021/14335-0, all granted by São Paulo Research Foundation (FAPESP), Brazil.

\appendix

\section{Overdamped limit for the heat in the active case}\label{A}
Considering the the most general case with both passive and active noises, we show here that the term for the change in kinetic energy appearing in the heat expression arises naturally if one starts from the underdamped equation of motion. In this regime, the velocity of the particle obeys
\begin{equation}
    m \dot{\boldsymbol{v}}(t) = -\gamma \boldsymbol{v}(t) + \boldsymbol{f}(t) +\boldsymbol{f}_a(t)+ \boldsymbol{\eta}(t). \label{underdamped}
\end{equation}
We defined the exchange of energy between the system and the passive thermal bath as
\begin{equation}
    Q[
\boldsymbol{v}] = \int_0^\tau \left(-\gamma (\boldsymbol{v}-\boldsymbol{f}_a\gamma^{-1})+ \boldsymbol{\eta}\right)\cdot (\boldsymbol{v}-\boldsymbol{f}_a\gamma^{-1})dt \label{heatunder}
\end{equation}
Plugging Eq.~\eqref{underdamped} into Eq.~\eqref{heatunder}, we find
\begin{equation}
    Q[
\boldsymbol{v}] = \int_0^{\tau} m \dot{\boldsymbol{v}} \cdot (\boldsymbol{v}-\boldsymbol{f}_a\gamma^{-1})dt- \int_0^{\tau} \boldsymbol{f}(t) \cdot (\boldsymbol{v}-\boldsymbol{f}_a\gamma^{-1})dt
\end{equation}
The first term in the integral can be rewritten as the change in kinetic energy since $\dot{\boldsymbol{v}} \cdot \boldsymbol{v} = \frac{d}{dt}(\boldsymbol{v}^2/2)$:
\begin{equation}
    Q[
\boldsymbol{v}] = \Delta K - \int_0^{\tau} \boldsymbol{f}(t)\cdot (\boldsymbol{v}-\boldsymbol{f}_a\gamma^{-1})dt + \frac{m}{\gamma} \int_0^\tau \dot{\boldsymbol{v}} \cdot \boldsymbol{f}_a dt.
\end{equation}
In the above expression we have the kinetic energy plus the work, but comparing with the overdamped case defined in Eq.~\eqref{activeheat}, we identify an extra term proportional to $\dot{\boldsymbol{v}} \cdot\boldsymbol{f}_a$. This kind of term can be hard to deal with as it corresponds to a product of two Gaussian stochastic processes. However, we can now appropriately take the overdamped limit where $m/\gamma$ is much smaller than the other timescales. By doing so, this new term can be disregarded while the kinetic energy term survives, giving
\begin{equation}
    Q[ \boldsymbol{v}] = \Delta K - \int_0^{\tau} \boldsymbol{f}(t) \cdot (\boldsymbol{v}-\boldsymbol{f}_a\gamma^{-1})dt.
\end{equation}
This is the same result that we obtained by starting from the overdamped equation of motion and defining heat including the kinetic energy by hand. Therefore, our expression is consistent with the underdamped case.

\section{Stationary distribution of velocities} \label{B}

The steady state distribution for the active case, given in Eq.~\eqref{steadyactive}, can be derived if one calculates the underdamped distribution of the velocity and take the overdamped limit. This means that the steady state distribution is a consequence of the overdamped regime. Since the underdamped distribution of the velocity is Gaussian, we just need to calculate its mean and the variance. 

The formal solution of the underdamped equation of motion, Eq.~\eqref{underdamped}, is
\begin{equation}
    \boldsymbol v_\tau = \boldsymbol v_0e^{-\frac{\gamma \tau}{m}} + \frac{e^{-\frac{\gamma \tau}{m}}}{m}\int_0^\tau e^{\frac{\gamma t'}{m}} (\boldsymbol f(t')+\boldsymbol \eta(t')+\boldsymbol f_a(t'))dt'
\end{equation}
The means is
\begin{equation}
    \langle \boldsymbol v_\tau\rangle = \langle \boldsymbol v_0 \rangle e^{-\frac{\gamma \tau}{m}} + \frac{e^{-\frac{\gamma \tau}{m}}}{m}\int_0^\tau e^{\frac{\gamma t'}{m}} \boldsymbol f(t') dt'
\end{equation}
In the overdamped limit, timescales are such that $\tau\gg m/\gamma$, giving
\begin{equation}
   \lim_{\tau_\gamma/\tau \rightarrow 0}\langle \boldsymbol v_\tau\rangle = \frac{\boldsymbol f(\tau)}{\gamma}.
\end{equation}
The variance of each component $i=x,y$ is the same and given by
\begin{eqnarray}
    \sigma^2_{v_\tau} = \sigma_{v_0}^2 e^{-2\frac{\gamma \tau}{m}} + \frac{e^{-2\frac{\gamma \tau}{m}}}{m^2}\int_0^\tau ds \int_0^\tau ds' e^{\frac{\gamma (s+s')}{m}}\times \nonumber \\
    \left[\langle \eta_i(s)  \eta_i(s')\rangle + \langle  f_{a,i}(s)  f_{a,i}(s')\rangle \right].
\end{eqnarray}
where a cross term obtained by squaring the two terms in the velocity solution is canceled out by the squared average velocity.
Solving both integrals by knowing the correlations of the noises, we find
\begin{eqnarray}
     \sigma^2_{v_\tau} = \left(\sigma_{v_0}^2 -\frac{1}{m\beta}  \right) e^{-2\tau/\tau_\gamma} + \frac{1}{m\beta} + \frac{v_a^2}{2\left(\frac{\tau_\gamma}{\tau_a}+1\right)} \nonumber \\+\frac{\tau_a v_a^2 \left(-(\tau_a + \tau_\gamma) e^{-2\tau/\tau_\gamma} + 2 \tau_a e^{-\tau \left(\frac{1}{\tau_\gamma} + \frac{1}{\tau_a}\right)} - \tau_a + \tau_\gamma \right)}{2 \left(\tau_\gamma^2 - \tau_a^2\right)}. \nonumber\\\label{b5}
\end{eqnarray}
Since in the overdamped regime $\tau\gg \tau_\gamma \equiv m/\gamma$, the terms with $e^{-2\tau/\tau_\gamma}$ vanish, and knowing that $\tau_a\gg \tau_\gamma$, the last term in Eq.~(\ref{b5}) vanishes, leading to
\begin{equation}
    \lim_{\tau_\gamma/\tau \rightarrow 0} \sigma^2_{v_\tau} = \frac{1}{m\beta}\left( 1+ \frac{\beta mv_a^2}{2\left({\tau_\gamma}/{\tau_a}+1\right)}\right). \label{eq:variancia}
\end{equation}
Therefore, the distribution in the overdamped limit becomes
\begin{eqnarray}
    P(\boldsymbol v_0) \sim \exp\left(\frac{-\frac{\beta m }{2}\left(\boldsymbol v_0- \boldsymbol f(0)/\gamma\right)^2}{\left( 1+ \frac{\beta mv_a^2}{2(\tau_\gamma/\tau_a+1)}\right)} \right),
\end{eqnarray}
where we used that $\langle \boldsymbol v_0\rangle = \boldsymbol f(0)/\gamma$ since we are writing the distribution of the initial velocities. For the distribution of final velocities, one just needs to use $\boldsymbol v_\tau$ and $\boldsymbol f(\tau)$.
Now, expanding the term $\beta m \left(\boldsymbol v_0- \boldsymbol f(0)/\gamma\right)^2/2$ in the argument of the exponential gives
\begin{equation}
    \frac{\beta m}{2} \boldsymbol v_0^2 - \beta \frac{m }{\gamma} \boldsymbol f(0) \cdot \boldsymbol v_0+ \beta\frac{m}{2\gamma^2}\boldsymbol f(0)^2.
\end{equation}
In the overdamped limit, we can ignore the two last terms and find
\begin{equation}
      P_a(\boldsymbol v_0) \sim \exp\left(-\frac{\beta m }{2}\boldsymbol v_0^2\left(1+\frac{\beta m v_a^2}{2(\tau_\gamma/\tau_a+1)}\right)^{-1}\right),
\end{equation}
which is the steady-state distribution of velocities for the AOUP model.

\section{Joint characteristic function}\label{jointApp}

The joint characteristic function can be calculated by noticing that the joint distribution of heat and work is
\begin{equation}
    P_{Q,W}(Q,W) = P_{\Delta K}(Q-W) P_W(W).\label{jointDIST}
\end{equation}
The characteristic function of each marginal distribution is
\begin{eqnarray}
    P_W(W) = \int e^{-i\lambda W} Z_W(\lambda)  d\lambda, \\P_{\Delta K } (Q-W) = \int e^{-i\lambda(Q-W)} Z_{\Delta K}(\lambda) d\lambda. 
\end{eqnarray}
At the same time, the joint characteristic function is related to the joint distribution by
\begin{equation}
    P_{Q,W}(Q,W) = \int \int e^{-i\lambda Q}e^{-i\lambda' W}Z_{Q,W}(\lambda, \lambda ') d\lambda d\lambda '.\label{eqseila}
\end{equation}
 Writing Eq.~\eqref{jointDIST} using the marginal characteristic, we have
 \begin{equation}
     P_{Q,W}(Q,W) = \int\!\!\! \int e^{-i\lambda (Q-W)}e^{-i\lambda ' W} Z_{\Delta K}(\lambda) Z_W(\lambda ') d\lambda d\lambda'. 
 \end{equation}
By making the transformation $\lambda'-\lambda=\lambda''$, we have
\begin{equation}
     P_{Q,W}(Q,W) = \int\!\!\! \int e^{-i\lambda Q} e^{-i\lambda '' W}Z_{\Delta K}(\lambda) Z_W(\lambda''+\lambda) d\lambda d\lambda ''.
\end{equation}
Therefore, by comparing with Eq.~\eqref{eqseila}, we can identify
\begin{equation}
    Z_{Q,W}(\lambda, \lambda ') = Z_{\Delta K}(\lambda) Z_W(\lambda'+\lambda).
\end{equation}

\section{Central moments via characteristic function}\label{C}
We can calculate all the order-$n$ central moments of the heat distribution via the characteristic function since
\begin{equation}
    \langle Q^n\rangle = i^n \frac{\partial^n Z_Q(\lambda)}{\partial \lambda^n}\bigg{|}_{\lambda \rightarrow 0},\label{Cfmeans}
\end{equation}
where $Z_Q(\lambda)$ is the characteristic function of the heat. This function can be constructed by multiplying the characteristic function of the kinetic energy, Eq.~\eqref{CFkineticenergy}, by that of the work, which is Gaussian, leading to
\begin{equation}
    Z_Q(\lambda) =  \frac{e^{-\frac{ f_0^2 m^2 \lambda^2 \left(\sigma_{v_0}^2 + \sigma_{v_\tau}^2\right)}{2 \gamma^2 \left(i + m \lambda \sigma^2_{v_0}\right) \left(-i + m \lambda \sigma_{v_\tau}^2\right)}} }{\left(1-i\lambda m\sigma_{v_0}^2\right)\left(1+i\lambda m\sigma_{v_\tau}^2\right)}e^{-\frac{1}{2} \lambda ^2 \sigma_W^2-i \lambda  \mu }.
\end{equation}
By using Eq.~\eqref{Cfmeans}, we can construct the mean, variance, skewness, and excess kurtosis.
The \textbf{mean} is
\begin{equation}
    \mu_Q = \mu_W +m \left(\sigma_{v_\tau}^2-\sigma_{v_0}^2\right).
\end{equation}
For an isothermal process, the heat mean is equal to the work mean. The \textbf{variance} is
\begin{align}
   \sigma_Q^2 = \sigma_{\Delta K}^2 + \sigma_W^2 
\end{align}
where
\begin{equation}
  \sigma_{\Delta K}^2 =  m^2 \left(\sigma_{v_0}^4 + \sigma_{v_\tau}^4 + \frac{f_0^2 (\sigma_{v_0}^2 + \sigma_{v_\tau}^2)}{\gamma^2}\right). \label{eq:sigmaK}
\end{equation}
The \textbf{skewness} measures the asymmetry of the distribution and depends only on the difference of the velocities variances, reading
\begin{equation}
    s_Q = 2 m^3 \left(\frac{3 f_0^2 (\sigma_{v_\tau}^4 - \sigma_{v_0}^4)}{\gamma^2} - \sigma_{v_0}^6 + \sigma_{v_\tau}^6\right).
\end{equation}
Finally, the \textbf{excess kurtosis} is defined as
\begin{equation}
    \kappa _Q = \frac{\langle \left(Q -\langle Q\rangle\right)^4\rangle}{\sigma_Q^4}-3,
\end{equation}
which, in terms of other variances, reads
\begin{equation}
     \kappa_Q = \frac{6 \gamma^2 m^4 \left(\gamma^2 \left(\sigma_{v_0}^8 + \sigma_{v_\tau}^8\right) + 4 f_0^2 \left(\sigma_{v_0}^6 + \sigma_{v_\tau}^6\right)\right)}{\left(2 f_0^2 m^2 \left(\sigma_{v_0}^2 + \sigma_{v_\tau}^2\right) + \gamma^2 \left(m^2 \left(\sigma_{v_0}^4 + \sigma_{v_\tau}^4\right) + \sigma_w^2\right)\right)^2}.
\end{equation}

\bibliography{references}

\end{document}